\documentclass[aps, superscriptaddress]{revtex4-1}

\pdfoutput=1
\usepackage{graphicx}

\usepackage{latexsym}
\usepackage[fleqn]{amsmath}
\usepackage{bm}
\usepackage{color}
\usepackage{amssymb}
\usepackage{bm}
\usepackage{ascmac}
\usepackage{bm}
\usepackage{comment}

\newcommand{\la}{\left\langle}
\newcommand{\ra}{\right\rangle}

\newcommand{\bra}[1]{\langle{#1}|}
\newcommand{\ket}[1]{|{#1}\rangle}

\newcommand{\Tr}{{\rm Tr}\hspace{0.07cm}}

\newcommand{\abs}[1]{{|#1|}}

\begin{document}
\title{Enhancement of quantum synchronization via continuous measurement and feedback control}
\author{Yuzuru Kato}
\email{Corresponding author: kato.y.bg@m.titech.ac.jp}
\affiliation{Department of Systems and Control Engineering,
	Tokyo Institute of Technology, Tokyo 152-8552, Japan}
\author{Hiroya Nakao}
\affiliation{Department of Systems and Control Engineering,
	Tokyo Institute of Technology, Tokyo 152-8552, Japan}
\date{\today}

\begin{abstract}
We study synchronization of a quantum van der Pol oscillator 
with a harmonic drive and demonstrate that 
quantum synchronization can be enhanced by performing continuous 
homodyne measurement on an additional bath linearly coupled to the oscillator and applying 
feedback control to the oscillator.
The phase coherence of the oscillator is increased by
reducing quantum fluctuations via the continuous measurement,
whereas the measurement backaction inevitably induces fluctuations
around the phase-locking point.
We propose a simple feedback policy for suppressing 
measurement-induced fluctuations 
by adjusting the frequency of the harmonic drive, 
which results in enhancement of quantum synchronization.
We further demonstrate that the maximum enhancement of quantum synchronization 
is achieved by performing quantum measurement on the quadrature angle at which the phase diffusion of the oscillator is the largest
and the maximal information on the oscillator phase is extracted.
\end{abstract}

\maketitle

\section{Introduction}
Studies pertaining to synchronization of nonlinear oscillators began 
in the 17th century when Huygens first documented the discovery of mutual synchronization 
between two pendulum clocks. 
Henceforth, synchronization phenomena have been widely observed
in various fields of science and technology, e.g., 
laser oscillations, chemical oscillations, 
spiking neurons, chorusing crickets, and mechanical vibrations
\cite{winfree2001geometry, kuramoto1984chemical, 
	pikovsky2001synchronization, nakao2016phase, strogatz1994nonlinear}.
Furthermore, synchronization have also been analyzed 
in engineering applications, 
such as voltage standards \cite{shapiro1963josephson}, injection locking \cite{adler1946study}, phase-locked loops in electrical circuits \cite{best1984phase},
and deep brain stimulation for the treatment of Parkinson's disease \cite{tass2001desynchronizing}.
Experimental studies of synchronizing nonlinear oscillators 
have recently reached the micrometer and nanometer scales,
including micro- and nanomechanical oscillators
\cite{shim2007synchronized,zhang2012synchronization,
zhang2015synchronization,bagheri2013photonic, matheny2014phase,
matheny2019exotic, colombano2019synchronization, liao2019quantum}, microlasers \cite{kreinberg2019mutual},
and spin-torque oscillators \cite{singh2019mutual},
and
experimental demonstrations of quantum phase synchronization in spin-$1$ atoms \cite{laskar2020observation}
and on the IBM Q system \cite{koppenhofer2020quantum} 
have been reported.
Owing to such experimental developments, 
the theoretical analysis of quantum synchronization
has received significant attention.
For example, synchronization of a quantum van der Pol (vdP) oscillator
with harmonic drive \cite{lee2013quantum, walter2014quantum, kato2019semiclassical, kato2020semiclassical, kato2020quantum, lorch2016genuine, weiss2017quantum} 
and squeezing \cite{sonar2018squeezing}, 
synchronization of coupled quantum vdP oscillators
\cite{lee2014entanglement, walter2015quantum, es2020synchronization, davis2018dynamics},
quantum spin systems \cite{roulet2018synchronizing, roulet2018quantum},
optomechanical systems \cite{ludwig2013quantum, weiss2016noise, amitai2017synchronization},
ensembles of atoms \cite{xu2014synchronization, xu2015conditional},
and isolated quantum systems \cite{witthaut2017classical},
quantum chimera states \cite{bastidas2015quantum},
quantum synchronization blockade \cite{lorch2017quantum},
and probing quantum synchronization using spin correlations \cite{hush2015spin}
have been studied theoretically.
These studies have revealed 
that quantum fluctuations generally induce phase diffusion in quantum limit-cycle oscillators and disturb strict synchronization
\cite{lee2013quantum, walter2014quantum, lee2014entanglement, sonar2018squeezing, kato2019semiclassical}.
To overcome this deleterious effect of quantum fluctuations on synchronization, 
Sonar {\em et~al.} studied the effect of squeezing and demonstrated that the entrainment of a quantum vdP oscillator to a squeezing signal can suppress quantum fluctuations, and consequently enhance quantum synchronization \cite{sonar2018squeezing}.

Measurement is one of the peculiar features in quantum systems;
it changes the quantum state of a system 
depending on the probabilistic outcomes
\cite{von1932mathematical, nielsen2000quantum}.
When knowledge about a system is indirectly obtained
by continuously monitoring the output of a field environment 
interacting  with an open quantum system,
the system dynamics under the measurement can be described by a
continuous quantum trajectory, 
i.e., a stochastic 
evolution of the system conditioned by the 
measurement outcomes \cite{belavkin1992quantum, wiseman2009quantum}.
This continuous measurement framework facilitates the  
investigation of novel dynamical features of quantum measurement,  
such as state preparation \cite{jacobs2010feedback, kato2014structure},
dynamical creation of entanglement \cite{gu2006entangled}, 
and unveiling \cite{bhattacharya2000continuous, scott2001quantum} 
and controlling \cite{eastman2019controlling} the chaotic 
behavior of quantum systems. 
It is also notable that
the experimental realization of continuous measurement
has been investigated recently \cite{hatridge2013quantum, minev2019catch, hays2020continuous}.

\begin{figure} [!t]
	\begin{center}
		\includegraphics[width=0.75\hsize,clip]{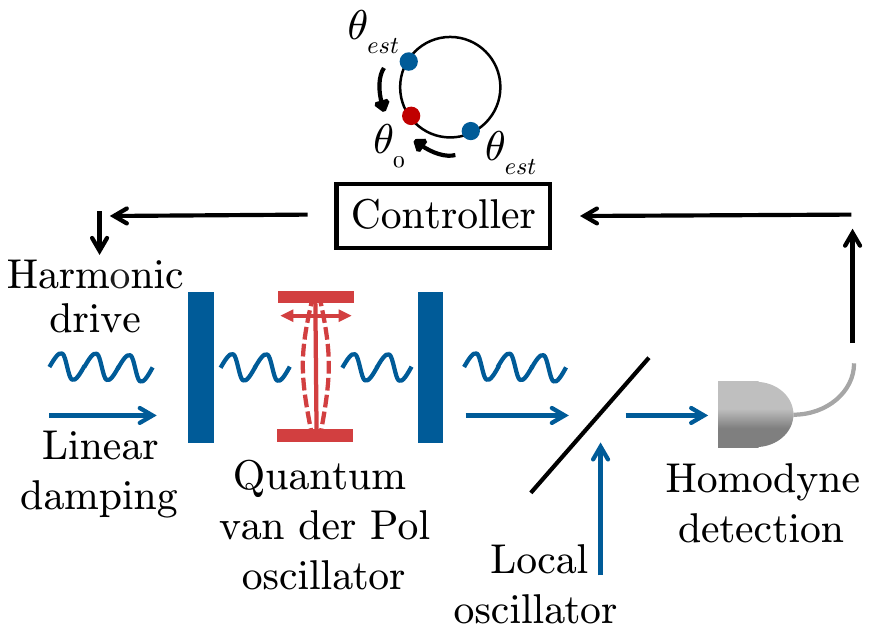}
		\caption{
			Enhancement of synchronization of a quantum vdP oscillator 
			with a harmonic drive via the continuous homodyne measurement and feedback control.
		}
		\label{fig_1}
	\end{center}
\end{figure}

Furthermore, the effect of continuous measurement of quantum limit-cycle oscillators 
has also been investigated, 
such as measurement-induced transitions between in-phase and anti-phase quantum synchronization \cite{weiss2016noise}, 
enhancement of nonclassicality 
in optomechanical oscillators 
via measurement
\cite{koppenhofer2018unraveling}, 
improvement in accuracy of 
Ramsey spectroscopy through 
measurement of synchronized atoms
\cite{xu2015conditional},
characterization of synchronization using quantum trajectories
\cite{es2020synchronization},
realization of quantum relaxation oscillators \cite{chia2020relaxation},
and instantaneous quantum phase synchronization of two decoupled quantum limit-cycle oscillators induced by conditional photon detection 
\cite{kato2020instantaneous}.
However, to the best of our knowledge, 
the effect of continuous measurement on
the enhancement of quantum synchronization has never been discussed.

In this study, 
we consider synchronization of a quantum vdP oscillator 
with a harmonic drive and demonstrate that 
performing continuous homodyne measurement on an additional bath linearly coupled to the oscillator and applying a feedback control 
to the oscillator can enhance quantum synchronization.
We demonstrate that quantum fluctuations disturbing the phase coherence 
can be reduced by continuous homodyne measurement,
and that the measurement backaction inevitably induces
stochastic deviations from the phase-locking point.
We propose a simple feedback policy 
that can suppress the fluctuations by 
adjusting the frequency of the harmonic drive.
Furthermore, we demonstrate that the maximal
enhancement of quantum synchronization 
can be achieved by performing the measurement on the quadrature angle 
at which the phase diffusion of the oscillator 
is the largest and the maximal information on 
the phase of the oscillator
can be obtained via the measurement.
It is shown that, by using continuous measurement and feedback control, 
more significant enhancement of phase coherence can be achieved than simply 
optimizing the waveform of the periodic amplitude modulation of the driving signal in the feed-forward setting as analyzed in our previous study~\cite{kato2020semiclassical}.
%

%
\section{Model}

We consider a quantum vdP oscillator~\cite{lee2013quantum, walter2014quantum} subjected to a harmonic drive.
A schematic diagram of the physical setup is shown in Fig.~\ref{fig_1}.
We further introduce an additional linear bath coupled to the oscillator,
perform continuous homodyne measurement of the output field from 
the oscillator to the bath,
and apply a feedback control to adjust the frequency of the harmonic drive 
(Fig.\ref{fig_1}).
The quantum vdP model is derived by considering a harmonic oscillator (main system) linearly and nonlinearly coupled to environmental baths and tracing out the environmental degrees of freedom. It is commonly assumed that baths have large degrees of freedom and the correlation times of the baths are significantly shorter than the timescale of the main system. The Markovian approximation of the baths can then be employed to derive the reduced master equation in the Lindblad form describing the main system under the effect of linear and nonlinear dissipation, yielding the quantum vdP model~\cite{gardiner1991quantum,lee2013quantum}.	

We denote by $\omega_{0}$ and $\omega_{d}$
the frequencies of the quantum vdP oscillator and harmonic drive, respectively.
The stochastic master equation
of the quantum system in the coordinate 
frame rotating with the frequency $\omega_{d}$ is written as 
\cite{lee2013quantum, walter2014quantum, kato2019semiclassical}
\begin{align}
\label{eq:sme}
d\rho = & \big\{
-i \left[  - (\Delta + \Delta_{fb}) a^{\dag}a 
+ i E(a - a^{\dag})
,\rho\right] 
+ \gamma_{1} \mathcal{D}[a^{\dag}]\rho + \gamma_{2}\mathcal{D}[a^{2}]\rho
\cr
& + \gamma_{3}\mathcal{D}[a]\rho \big\} dt
+ \sqrt{\eta \gamma_{3}}\mathcal{H}[a e^{- i \theta}]\rho dW,
\cr
dY
=& \sqrt{\eta \gamma_{3}}\hspace{0.05cm}{\rm Tr}[(a e^{- i \theta} 
+ a^{\dag} e^{ i \theta})\rho]dt 
+ dW,
\end{align}
with
$
\mathcal{D}[L]\rho
=L \rho L^{\dag}-\frac{1}{2}(L^{\dag}L\rho + \rho L^{\dag} L),
\mathcal{H}[L]\rho = L \rho + \rho L^{\dag}
- \Tr[(L + L^\dagger) \rho]\rho,$
where $\mathcal{D}$ is the Lindblad form and $\mathcal{H}[a e^{- i \theta}]$ characterizes the measurement on the quadrature $a e^{-i \theta} + a^{\dag} e^{i \theta}$,
$\rho$ is the density matrix representing the system state, 
$a$ and $a^{\dag}$ denote the annihilation and creation operators
($\dag$ represents Hermitian conjugate), respectively,
$\Delta = \omega_{d} - \omega_{0}$ is the frequency detuning of
the harmonic drive from the oscillator, 
$\Delta_{fb}$ is the feedback control to adjust the frequency detuning, i.e.,
the frequency of the harmonic drive, 
$E$ is the intensity of the harmonic drive, 
$\gamma_{1}$, $\gamma_{2}$, and $\gamma_{3}$ represent the decay 
rates for the negative damping, nonlinear damping, and linear damping, respectively,
$\eta$ is the efficiency of the measurement
(we set $\eta = 1$ when the measurement is performed, and $\eta = 0$ 
when it is not),
$\theta$ specifies the quadrature angle of the measurement,
$W$ represents a Wiener process satisfying 
$\mathbb{E}[dW] = 0$ 
and
$\mathbb{E}[dW^2] = dt$,
$Y$ is the output of the measurement result, 
and the reduced Planck's constant is set as $\hbar = 1$.

In the following, we use the parameter settings such that  
the oscillator is synchronized with the harmonic drive and the Wigner distribution,
a quasiprobability distribution \cite{gardiner1991quantum}, of the steady-state density matrix $\rho_{ss}$ of Eq.~(\ref{eq:sme}) 
is concentrated
around a stable phase-locking point along the limit-cycle orbit in the classical limit (see, e.g., Fig.~\ref{fig_3}(a)) when the measurement is not performed ($\eta=0$).

We set the feedback control $\Delta_{fb}$ as (see Appendix~\ref{appendix_a} for details)
\begin{align}
\label{eq:fb}
\Delta_{fb} = -K_{fb}(\theta_{est} - \theta_0),
\end{align}
where $K_{fb}~(>0)$ represents the feedback gain
and $\theta_{0} = \arctan \left( 
{\Tr[p \rho_{ss}] / \Tr[x \rho_{ss}]} \right)$ 
represents the locking phase in the absence of the measurement,
which is calculated as the angle between the expectation values
of the position operator $x = (a + a^{\dag})/2$
and the momentum operator $p = -i (a - a^{\dag})/2$ with respect to 
the steady-state $\rho_{ss}$ of Eq.~(\ref{eq:sme}) without measurement,
and $\theta_{est} = \arctan \left( \Tr[p \rho_{est}] / \Tr[x \rho_{est}] \right)$, which is chosen such that $- \pi + \theta_0 \leq \theta_{est} < \theta_0 + \pi$, represents the phase of the system 
calculated from the instantaneous state $\rho_{est}$ of Eq.~(\ref{eq:sme}) with measurement, which is conditioned on the measurement record.
The feedback control above can actually suppress the fluctuations of the system state around the phase-locking point as will be shown 
in the next section.

Note that we cannot introduce the feedback control without the linear bath in the present scheme because the continuous measurement should be performed on the linear bath coupled to the main system.
Because we introduce the feedback control in the frequency detuning, the instantaneous frequency of the driving signal varies (when the natural frequency of the vdP oscillator is fixed). It is assumed that 
this variation in the driving frequency due to the feedback control is much smaller than the frequencies $\omega_0$ and $\omega_d$ and, when the feedback control works and the synchronized state is maintained, the driving frequency remains approximately constant on average and the vdP oscillator is entrained to this frequency.

To evaluate the 
phase coherence of the quantum vdP oscillator, 
we use the order parameter
\cite{weiss2016noise, lorch2016genuine} 
\begin{align}
S_1  = \abs{S_1} e^{i\phi_1} = \frac{ \Tr[a \rho] }{\sqrt{\Tr[a^{\dag}a \rho]}},
\end{align}
which is a quantum analog of the order parameter for a single classical noisy oscillator \cite{kuramoto1984chemical,pikovsky2001synchronization}.
The absolute value $\abs{S_1}$ quantifies the degree of phase coherence 
and assumes the values in $0 \leq \abs{S_1} \leq 1$, 
where $\abs{S_1}=1$ when 
the oscillator state is perfectly 
phase-coherent and $\abs{S_1}=0$ 
when the state is perfectly phase-incoherent.
Here $\phi_1$ represents the average phase value of the oscillator.
%


\section{Results}
\label{sec3}
Numerical simulations of Eq.~(\ref{eq:sme}) are performed.
In Sections~\ref{sec3a},~\ref{sec3b}, and ~\ref{sec3c}, 
we set the parameter values in the semiclassical regime,
$(\Delta, \gamma_{2}, \gamma_{3}, E)/\gamma_{1} = 
(0.05, 0.05, 0.1, \sqrt{0.1})$, 
to clarify the relation between the quantum system and 
its classical limit \cite{kato2019semiclassical},
and the feedback gain is set as $K_{fb}/\gamma_{1} = 1$ when we
apply the feedback control. 
In Secs.~\ref{sec3d}, 
we discuss the applicability of the proposed scheme in the quantum regime
with parameter values
$(\Delta, \gamma_{2}, \gamma_{3}, E)/\gamma_{1} = 
(0.05, 0.25, 0.25, \sqrt{0.1})$,
and apply feedback control with a feedback gain $K_{fb}/\gamma_{1} = 7.5$.
In Sections~\ref{sec3a}, \ref{sec3b}, and \ref{sec3d}, we set $\theta = 0$ for
the quadrature of measurement and, in Section~\ref{sec3c}, the effect of varying $\theta$ is analyzed.
We always set the initial state of the simulation as the vacuum state, i.e., $\rho = \ket{0}\bra{0}$.

\subsection{Without feedback control}
\label{sec3a}
\begin{figure} [!t]
	\begin{center}
		\includegraphics[width=0.9\hsize,clip]{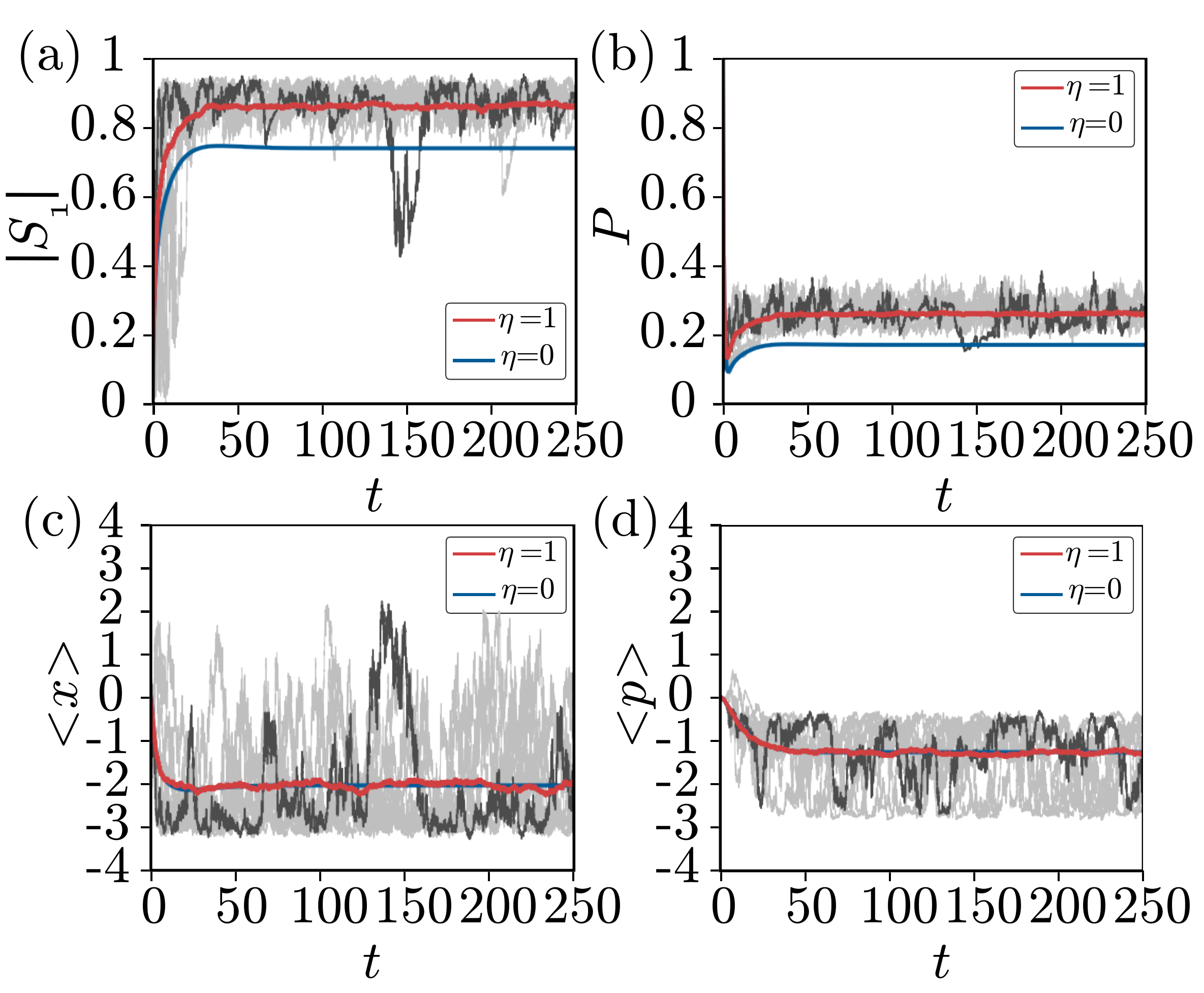}
		\caption{
			Measurement-induced increase in 
			phase coherence without feedback control in semiclassical regime.
			(a) Order parameter $|S_1|$.
			(b) Purity $P$.
			(c) Expectation values of position operator $\langle {x} \rangle$.
			(d) Expectation values of momentum operator $\langle {p} \rangle$.
			For the case with measurement ($\eta=1$), average values of results calculated from
			$300$ trajectories are shown by red lines and $10$ out of $300$ 
			individual trajectories are shown by gray lines, with the dark one representing a single realization of the trajectory.
			For the case without measurement ($\eta=0$), the results of a single trajectory are shown by blue lines.			
		}
		\label{fig_2}
	\end{center}
\end{figure}
\begin{figure} [!t]
	\begin{center}
		\includegraphics[width=0.8\hsize,clip]{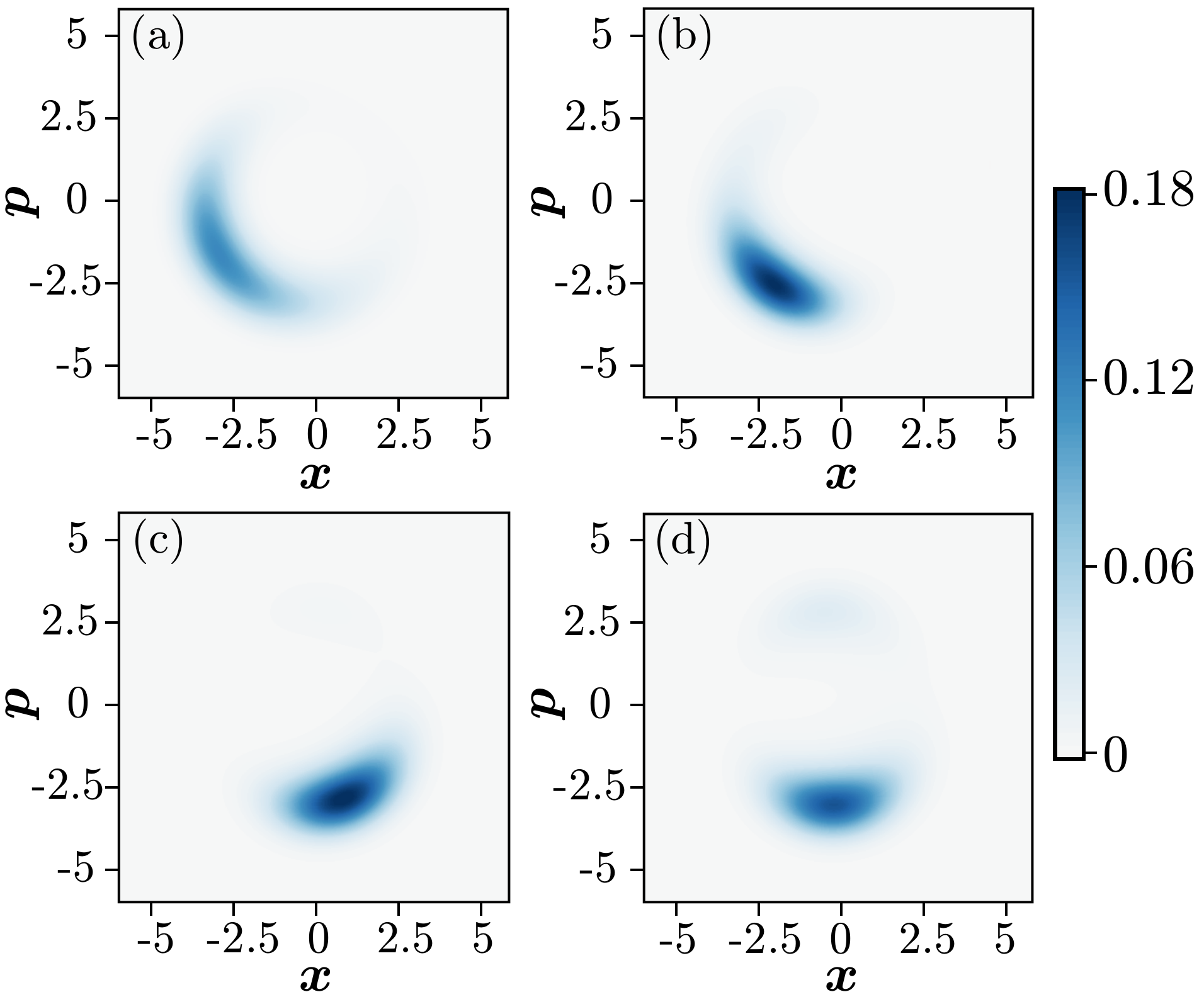}
		\caption{
			Wigner distributions of system without feedback control
			in semiclassical regime.
			(a) Wigner distribution of steady state of Eq.~(\ref{eq:sme}) without measurement.
			(b,c,d) Wigner distributions of three different trajectories of Eq.~(\ref{eq:sme}) with measurement at $t = 250$.
		}
		\label{fig_3}
	\end{center}
\end{figure}
We first consider the case without feedback control, i.e., $K_{fb} = 0$,
in the semiclassical regime.
When the measurement is performed, 
we calculated the average values over $300$ trajectories obtained by the numerical simulations of Eq.~(\ref{eq:sme}) from the same initial state ($\rho =\ket{0}\bra{0}$) 
because the system trajectories behaved stochastically.
The average results are compared with the results in the case without measurement when the system trajectory of Eq.~(\ref{eq:sme}) is deterministic.

Figures~\ref{fig_2}(a),~\ref{fig_2}(b),~\ref{fig_2}(c), and~\ref{fig_2}(d) 
show the trajectories of the absolute values of the order parameter $\abs{S_1}$
quantifying the degree of phase coherence,
the purity $P = \Tr[\rho^2]$, 
the expectation value of the position operator $\la x \ra = \Tr[x \rho]$, and 
the expectation value of the momentum operator $\la p \ra = \Tr[p \rho]$, respectively.
Note that these expectation values are fluctuating 
in the case with measurement.

As shown in Fig.~\ref{fig_2}(a), 
the average value of the order parameter $\abs{S_1}$ with measurement
is larger than the (deterministic) value of $\abs{S_1}$ without measurement,
e.g., $\abs{S_1} = 0.859$ with measurement and 
$\abs{S_1} = 0.737$ without measurement at $t = 250$,
signifying  
the phase coherence
increased on average due to the continuous homodyne measurement.
The increase in the purity is evident in Fig.~\ref{fig_2}(b); the average values of the purity $P$ with measurement are larger than the stationary value
of $P$ without measurement, e.g., $P = 0.258$ with measurement 
and $P = 0.169$ without measurement at $t = 250$ sufficiently after the initial relaxation. 

We note that the observed increase in $\abs{S_1}$ or $P$ is an average effect;
the values of these quantities for a single trajectory of Eq.~(\ref{eq:sme}) with the measurement fluctuates significantly 
and occasionally take smaller values than those without measurement, as shown by the dark gray lines in Figs.~\ref{fig_2}(a) and \ref{fig_2}(b).
We also note that 
the increase in the purity implies 
the reduction in the phase diffusion of the oscillator
(See Appendix~\ref{appendix_b}).
Owing to the increase in phase coherence by the measurement,
the measurement backaction inevitably induces fluctuations 
in the system state around the phase-locking point. 
It is evident from Figs.~\ref{fig_2}(c) and~\ref{fig_2}(d)
that $10$ trajectories of $\la x \ra$ and $\la p \ra$ 
obtained by simulating Eq.~(\ref{eq:sme}) with measurement (gray lines)
exhibit strong fluctuations based on the measurement outcomes.

The increase in phase coherence by the measurement 
is also observed in the Wigner distribution.
Figure~\ref{fig_3}(a) shows the steady-state Wigner distribution obtained from Eq.~(\ref{eq:sme}) without measurement
($\rho$ converges to a steady state in this case), and Figs.~\ref{fig_3}(b),~\ref{fig_3}(c), and~\ref{fig_3}(d) show
the instantaneous Wigner distributions at $t = 250$ of 
three trajectories obtained by simulating Eq.~(\ref{eq:sme}) with measurement ($\rho$ behaves stochastically in this case). 

Comparing Figs.~\ref{fig_3}(b),~\ref{fig_3}(c), and~\ref{fig_3}(d) with Fig.~\ref{fig_3}(a), 
increase in phase coherence by the continuous homodyne measurement
is observed from the strongly concentrated Wigner distributions.
We also observe that the location of the distribution differs 
by trajectory because the measurement backaction 
randomly disturbs the system state based on the measurement outcomes.
We here note that the increase in the phase coherence
of the Wigner distribution occurs for each stochastic realization of the stochastic master equation~(\ref{eq:sme}) with measurement.
If we average the Wigner distributions with measurement over all realizations, the result without measurement as shown in Fig.~\ref{fig_3}(a) is obtained.

\begin{figure} [htbp]
	\begin{center}
		\includegraphics[width=0.9\hsize,clip]{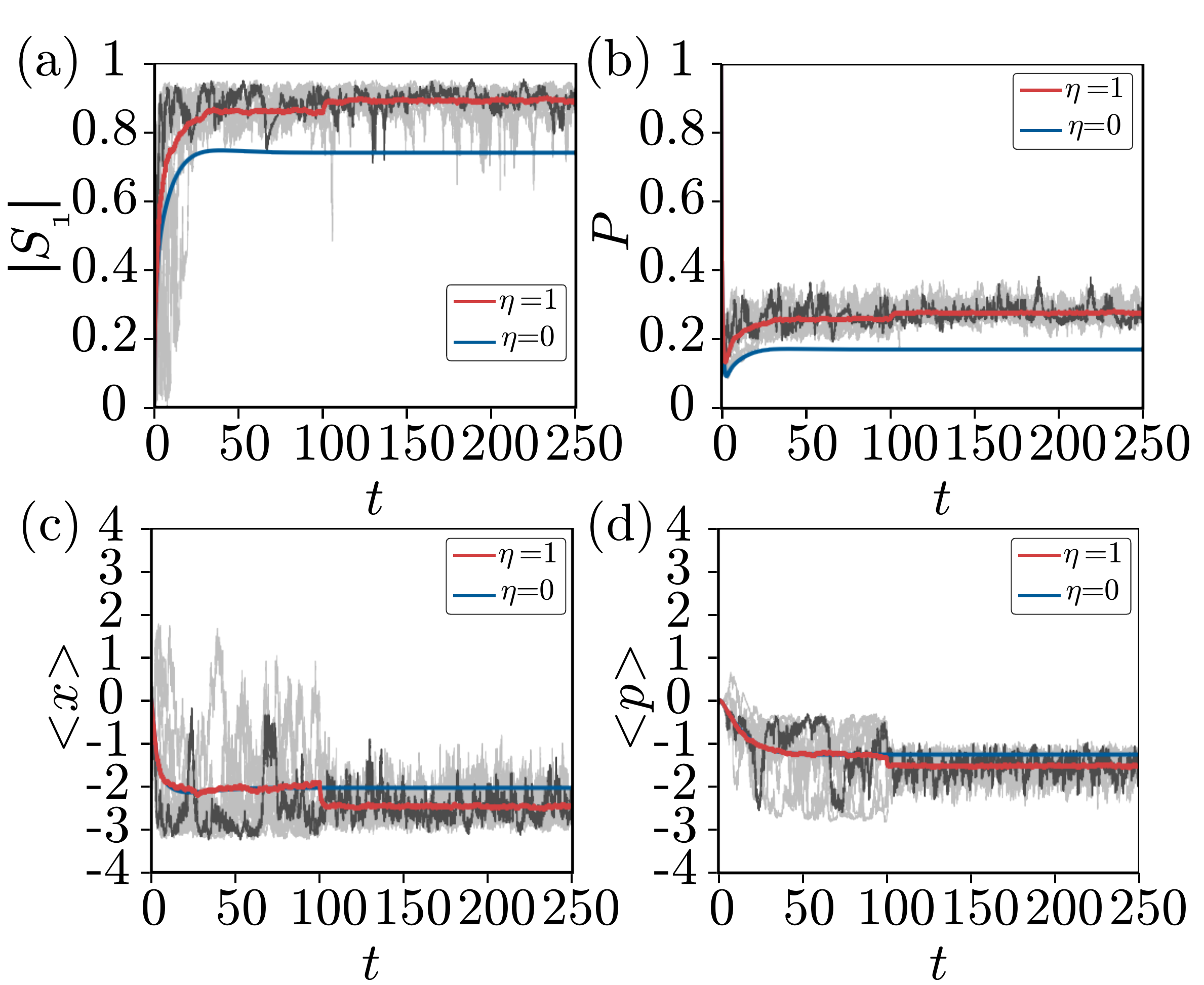}
		\caption{
			Measurement-induced enhancement of quantum synchronization with feedback control in semiclassical regime.
			Feedback control is applied from $t=100$.
			(a) Order parameter $|S_1|$.
			(b) Purity $P$.
			(c) Expectation values of the position operator $\langle {x} \rangle$.
			(d) Expectation values of the momentum operator $\langle {p} \rangle$.
			For the case with measurement ($\eta=1$), average values of results calculated from
			$300$ trajectories are shown by red lines and $10$ out of $300$ 
			individual trajectories are shown by gray lines with the dark one representing a single realization of the trajectory.
			For the case without measurement ($\eta=0$), the results of a single trajectory are shown by blue lines.			
		}
		\label{fig_4}
	\end{center}
\end{figure}

\begin{figure} [!t]
	\begin{center}
		\includegraphics[width=0.8\hsize,clip]{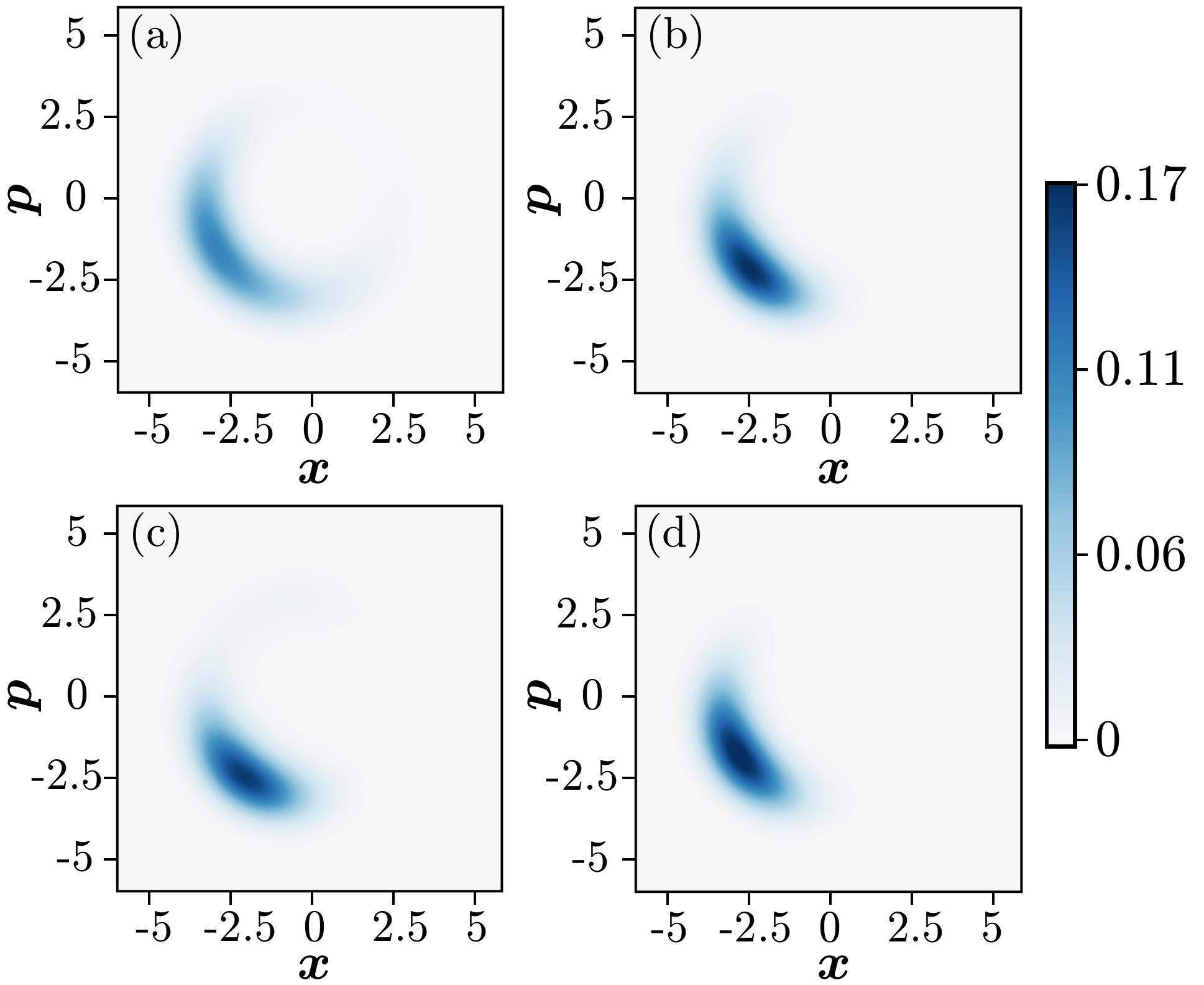}
		\caption{
			Wigner distributions of system with feedback control
			in semiclassical regime.
			(a) Wigner distribution of steady state of Eq.~(\ref{eq:sme}) without measurement.
			(b,c,d) Wigner distributions of three different trajectories of Eq.~(\ref{eq:sme}) with measurement at $t = 250$.
			Feedback control is applied from time $t=100$.
		}
		\label{fig_5}
	\end{center}
\end{figure}

\subsection{With feedback control}
\label{sec3b}

As presented in Section~\ref{sec3a}, we observed that the measurement  
increases phase coherence but induces fluctuations in the system state 
around the phase-locking point simultaneously.
To suppress the fluctuations of the system state,
we introduce the feedback control expressed in Eq.~(\ref{eq:fb}).

Figures~\ref{fig_4}(a),~\ref{fig_4}(b),~\ref{fig_4}(c), and~\ref{fig_4}(d) 
show the trajectories of $\abs{S_1}$, $P$, $\la x \ra$, and $\la p \ra$, respectively.
The feedback control is applied from $t=100$.
As shown in Fig.~\ref{fig_4}(a), 
the average order parameter $\abs{S_1}$ with measurement 
takes larger values than $\abs{S_1}$ without measurement,
e.g., $\abs{S_1} = 0.889$ with measurement and 
$\abs{S_1} = 0.737$ without measurement at $t = 250$.
We also see in Fig.~\ref{fig_4}(b) that the average values of 
$P$ with measurement are larger than those without
measurement, e.g., $P = 0.275$ with measurement 
and $P = 0.169$ without measurement at $t = 250$. 

The role of the feedback control is evident from Figs.~\ref{fig_4}(c) and~\ref{fig_4}(d), where $10$ trajectories of $\la x \ra$ and $\la p \ra$ obtained by simulating Eq.~(\ref{eq:sme}) with measurement are shown (gray lines).
The fluctuations around the phase-locking point are suppressed 
by the feedback control that is turned on after $t=100$.
We note that we used the same
sequences of the Wiener increments
in the numerical simulations of Eq.~(\ref{eq:sme}) 
in the case without feedback control.

The average values of $\la x \ra$ and $\la p \ra$ (red lines) are smaller than those for the case without measurement (blue lines). This can be explained as follows.
The backaction induces strong fluctuations in $\la x \ra$.
Without feedback control, $\la x \ra$, which fluctuates near the phase-locking point, occasionally exhibits a large increase along the limit-cycle trajectory 
to the clockwise direction.
This large increase in $\la x \ra$ is suppressed by the feedback control, which results in a smaller average value of $\la x \ra$.
Although the backaction is weaker for $\la p \ra$, the suppression of 
large increase in $\la x \ra$ by the feedback control results in a smaller average value of $\la p \ra$.
The effect of feedback control for suppressing the fluctuations 
of the system state is also evident from the Wigner distribution of the system.
Figure~\ref{fig_5}(a) shows the steady-state Wigner distribution of Eq.~(\ref{eq:sme}) without measurement, whereas Figs.~\ref{fig_5}(b), \ref{fig_5}(c), and \ref{fig_5}(d) show three realizations of the Wigner distributions at $t = 250$ of Eq.~(\ref{eq:sme}) with measurement.
Comparing Figs.~\ref{fig_5}(b),~\ref{fig_5}(c), and \ref{fig_5}(d) with
Figs.~\ref{fig_3}(b),~\ref{fig_3}(c), and \ref{fig_3}(d),
it is clear that the fluctuations around the phase-locking point  
are suppressed effectively by the feedback control.

The results above indicate that enhancement of synchronization,
i.e., larger phase coherence and smaller fluctuations around the phase-locking point,
can be achieved via continuous measurement and feedback control.

\subsection{Dependence on measurement quadrature}
\label{sec3c}

Thus far, we have fixed $\theta$,
the quadrature  of the measurement, at $0$. 
Next, we consider the effect of varying $\theta$
on the enhancement of quantum synchronization. 

\begin{figure} [!t]
	\begin{center}
		\includegraphics[width=0.8\hsize,clip]{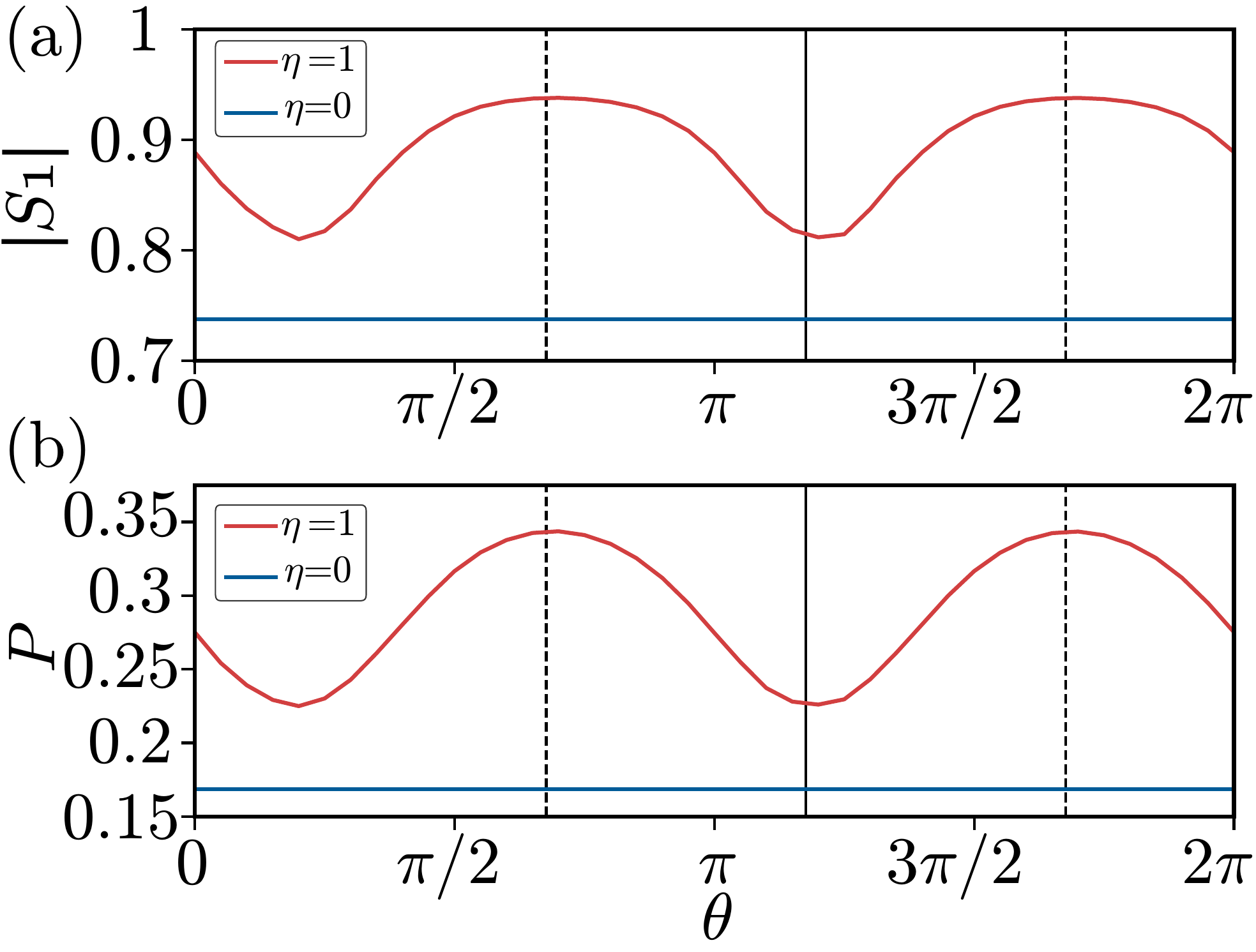}
		\caption{
			Dependence of results
			on measurement quadrature.
			(a) Order parameter $\abs{S_1}$.
			(b) Purity $P$.
			Order parameter and purity averaged over 
			$400$ trajectories with measurement at $t=250$ (red lines)
			are compared with those for a single trajectory without measurement (blue lines).
			Phase-locking point $\theta_0$ (a solid black vertical line)
			and points orthogonal to the phase-locking point $\theta_0 + \pi/2$ 
			(dotted black vertical lines) are shown.
		}
		\label{fig_6}
	\end{center}
\end{figure}

Figures~\ref{fig_6}(a) and~\ref{fig_6}(b) show the average values of $\abs{S_1}$ and $P$ at $t=250$ for $0 \leq \theta \leq 2\pi$, respectively, which are calculated from $400$ trajectories of Eq.~(\ref{eq:sme}) with measurement and 
the feedback control.
For comparison, we also show the values of $\abs{S_1}$ and $P$ for the steady state of Eq.~(\ref{eq:sme}) without the measurement.
The maximum values of $\abs{S_1}$ and $P$ are achieved at $\theta = 2.199$, which is 
approximately orthogonal to 
the locking phase, i.e., $\theta_0 = 3.696$. 

The result above can be interpreted as follows. 
The phase diffusion of 
the oscillator is maximized when 
$\theta$ is orthogonal to $\theta_0$,
and performing the measurement on the quadrature 
specified by this $\theta$
extracts the maximum information on the oscillator phase.
Hence, the maximum reduction in quantum fluctuations and
enhancement in synchronization are attained at the quadrature angle.
Note that uncertainty in the oscillator phase always exists due to quantum fluctuations and no matter which quadrature angle we choose, we can obtain non-zero information about the oscillator phase and enhance quantum synchronization via the continuous measurement.

\subsection{Applicability in stronger quantum regime}
\label{sec3d}

\begin{figure} [!t]
	\begin{center}
		\includegraphics[width=0.9\hsize,clip]{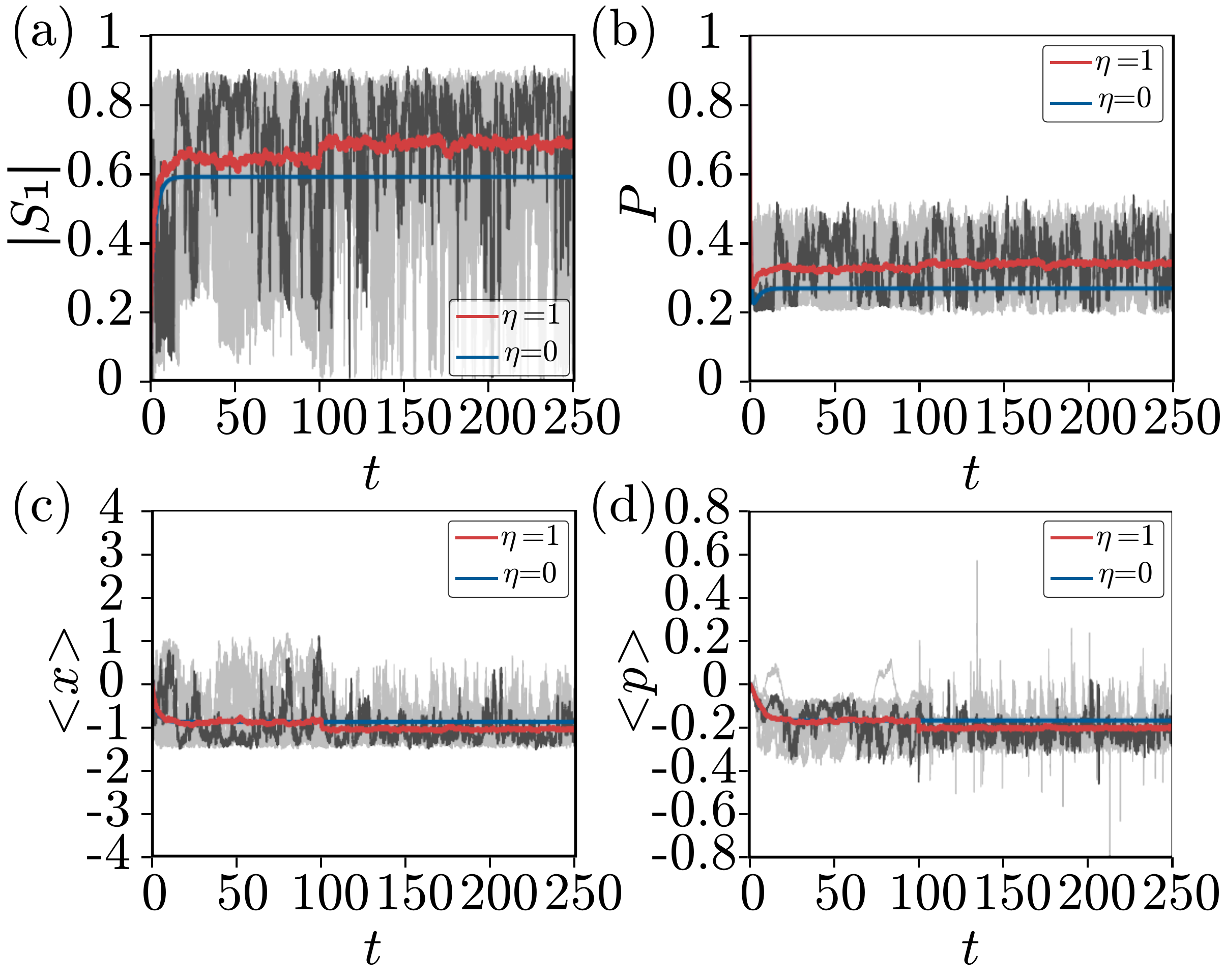}
		\caption{
			Measurement-induced enhancement of quantum synchronization with feedback control in quantum regime.
			Feedback control is applied from $t=100$.
			(a) Order parameter $|S_1|$.
			(b) Purity $P$.
			(c) Expectation values of position operator $\langle {x} \rangle$.
			(d) Expectation values of momentum operator $\langle {p} \rangle$.
			For the case with measurement ($\eta=1$), averaged values of results calculated from
			$300$ trajectories are shown by red lines, and $10$ out of $300$ individual trajectories are shown by gray lines with the dark one representing a single realization of the trajectory.
			For the case without measurement ($\eta=0$), results of a single trajectory are shown by blue lines.			
		}
		\label{fig_7}
	\end{center}
\end{figure}
\begin{figure} [!t]
	\begin{center}
		\includegraphics[width=0.8\hsize,clip]{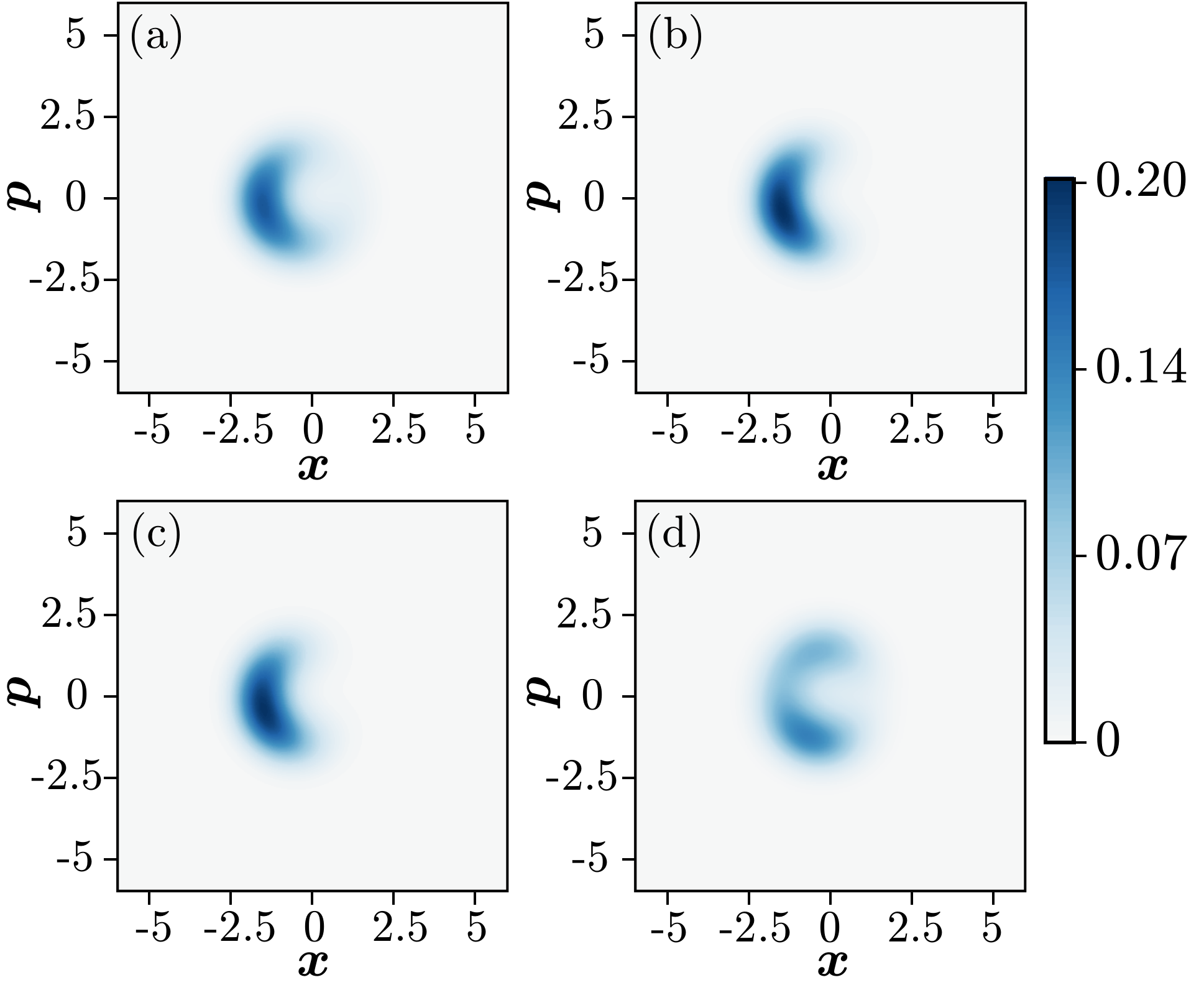}
		\caption{
			Wigner distributions of system with feedback control
			in quantum regime.
			(a) Wigner distribution of steady state of Eq.~(\ref{eq:sme}) without measurement.
			(b,c,d) Wigner distributions of three different trajectories of Eq.~(\ref{eq:sme}) with measurement at $t = 250$.
			Feedback control is applied from time $t=100$.			
		}
		\label{fig_8}
	\end{center}
\end{figure}

Finally, we discuss the enhancement of quantum synchronization 
via continuous measurement and feedback control in a stronger quantum regime.
The parameters are shown at the beginning of Section \ref{sec3}.
Figures~\ref{fig_7}(a),~\ref{fig_7}(b),~\ref{fig_7}(c), and~\ref{fig_7}(d) 
show the trajectories of $\abs{S_1}$, $P$, $\la x \ra$, and $\la p \ra$, respectively.
The feedback control is applied after $t=100$.

As shown in Fig.~\ref{fig_7}(a), 
the average order parameter $\abs{S_1}$ with measurement 
takes larger values than $\abs{S_1}$ without measurement,
e.g., $\abs{S_1} = 0.683$ with measurement and 
$\abs{S_1} = 0.586$ without measurement at $t = 250$.
We also see in Fig.~\ref{fig_7}(b) that the averaged values of 
$P$ with measurement are larger than the values of $P$ without
measurement, e.g., $P = 0.338$ with measurement 
and $P = 0.266$ without measurement at $t = 250$. 

As shown from the results above, both $\abs{S_1}$ and $P$ increase on average with measurement even in this quantum regime.
The suppression of the measurement-induced fluctuations 
by the feedback control is shown in Fig.~\ref{fig_7}(c),
where $10$ trajectories of $\la x \ra$ obtained by simulating 
Eq.~(\ref{eq:sme}) with measurement are shown (gray lines).
We see that the fluctuations in $\la x \ra$ around the phase-locking point 
are suppressed by the feedback control which is turned on after $t=100$.
However, 
in Fig.~\ref{fig_7}(d)
where $10$ trajectories of $\la p \ra$ obtained by simulating 
Eq.~(\ref{eq:sme}) with measurement are shown (gray lines),
the fluctuations in $\la p \ra$ still remain 
and can be even stronger after the feedback control is turned on at $t=100$.
We note that the fluctuations in $\la p \ra$ become smaller on average but $\la p \ra$ also exhibits occasional bursty increases when the feedback control is applied.
This is because the feedback control induces more localized states with stronger phase coherence than the case without feedback, and measurement-induced fluctuations of such states yield larger variations in $\la p \ra$.

These results are also observed from the Wigner distribution of the system.
Figure~\ref{fig_8}(a) shows the steady-state Wigner distribution of Eq.~(\ref{eq:sme}) without measurement and Figs.~\ref{fig_8}(b), \ref{fig_8}(c), and \ref{fig_8}(d) show three realizations of the Wigner distributions at $t = 250$ of Eq.~(\ref{eq:sme}) with measurement, respectively.
The fluctuations around the phase-locking point are suppressed effectively by the feedback control and the enhancement of quantum synchronization is achieved
in Figs.~\ref{fig_8}(b) and ~\ref{fig_8}(c).
However, the measurement-induced fluctuation still remains and the system occasionally exhibits transient short-time failures of synchronization with low phase coherence as shown in Fig.~\ref{fig_8}(d).

The numerical results shown in Figs.~\ref{fig_7} and~\ref{fig_8} indicate that quantum synchronization is enhanced only probabilistically in the strong quantum regime considered here.  
We repeated numerical simulations from the same initial condition and empirically 
obtained a probability of success approximately 80 percents for the enhancement of quantum synchronization, namely, Wigner distributions at time $t = 250$ are strongly localized around the phase-locking point $\theta_0$ for approximately 80 percent of the trajectories.
Thus, in this regime, because of the strong quantum fluctuations, the feedback control occasionally fails to suppress the measurement-induced fluctuations and enhance quantum synchronization.
We note here that, because the feedback signal is different for each trajectory,  the Wigner distribution is not averaged over trajectories but its individual realizations for different trajectories are plotted in Figs.~\ref{fig_5} and~\ref{fig_8}.

We also note that the strong quantum fluctuations lead to the weaker enhancement of synchronization.
This is evident in the improvement of $\abs{S_1} = 0.683$ 
from $\abs{S_1} = 0.586$ by a factor $0.683/0.586 = 1.166$
in the quantum regime, which is smaller than
the improvement of $\abs{S_1} = 0.889$ from 
$\abs{S_1} = 0.737$ by a factor $ 0.889/0.737 = 1.206$
in the semiclassical regime in Figs.~\ref{fig_4} and \ref{fig_5}.
More detailed and systematic numerical analysis in the strong quantum regime and finding an improved feedback strategy for achieving the 
enhancement of quantum synchronization 
to minimize the probability of failure as shown in Fig.~\ref{fig_8}(d) are the subject of future study.
%
%
%

\section{Discussion}	
We have confirmed that our scheme can enhance synchronization of the vdP oscillator to the harmonic drive. 
Here we discuss several perspectives for future studies.

(i)
Though not treated in this study, the effect of thermal noise~\cite{gardiner1991quantum} and measurement inefficiency~\cite{wiseman2009quantum} should be taken into account
to consider more realistic experimental situations.
If we consider these effects, the intensities of the feedback signal and the noise term in the model will change. However,
it is expected that, as long as their effects are not too strong, the qualitative behavior of the system will not change and enhancement of synchronization will occur because the essential mechanism of the scheme explained in the Appendices remains unchanged.

(ii)
We focused only on the synchronized regime and tried to improve the coherence by using continuous measurement and feedback. The effect of continuous measurement on the overall synchronization property in a wider frequency range, in particular, Arnold's tongue \cite{lorch2016genuine, lee2014entanglement}, is of great importance. For example, it is expected that the shape of Arnold's tongue can considerably change depending on the measurement inefficiency and will be studied in detail in our future work.

(iii)
In this study, we introduced the feedback control in the frequency of the external drive for suppressing measurement-induced fluctuations. 
Alternatively, we may adjust the quadrature of the harmonic drive based on the measurement outcomes and enhancing synchronization without changing the frequency of the external drive.
The well-known Markovian feedback \cite{wiseman2009quantum} may also be useful for enhancing quantum synchronization.
Apart from the feedback strategy, an additional linear-damping bath without measurement may also enhance synchronization of a quantum vdP oscillator with an external drive
in the deep quantum regime 
as discussed in Ref.~\cite{mok2020synchronization}.

(iv) It is interesting to investigate the connection between enhancement of phase coherence by measurement and uncertainty relation. The measurement plays two conflicting roles on synchronization: (i) enhancing the coherence by information gain (the term $\mathcal{H}[ae^{-i\theta}]$ in Eq.~(\ref{eq:sme})) and (ii) 
destroying the locking to the harmonic forcing by the disturbance (the term $\mathcal{D}[a]$ in Eq.~(\ref{eq:sme})) \cite{jacobs2006straightforward}. 
The dependence of the result on the optimal quadrature angle of the measurement is also related to the uncertainty principle.
A more detailed analysis on this point would provide insights into how the uncertainty relation affects the enhancement of quantum synchronization.

(v) It would also be important to perform a systematic analysis of the transition from the semiclassical regime to the quantum regime and analyze the performance of our feedback control in enhancing synchronization, which will clarify the applicability of our scheme in the strong quantum regime.

(vi) Studying the enhancement of quantum synchronization of a quantum vdP oscillator with squeezing drive \cite{sonar2018squeezing} by using the present scheme would also be interesting, because it is expected that the combining effects of enhancement of quantum synchronization via squeezing lights and our scheme of continuous measurement and feedback control would be useful for further enhancement of quantum synchronization.

\section{Conclusion}
We considered synchronization of a quantum van der Pol oscillator 
with a harmonic drive.
We demonstrated that introducing an additional linear bath
coupled to the system and performing continuous homodyne measurement
of the bath can increase the phase coherence of the system.
We also proposed a simple feedback policy for suppressing the fluctuations in the system state around the phase-locking point by adjusting the frequency of the harmonic drive,
and achieved the measurement-induced enhancement of synchronization.
We further demonstrated 
that the maximum enhancement of synchronization is achieved
when we perform measurement on the quadrature angle at which the phase diffusion of the oscillator is maximized and the maximum information regarding the oscillator phase is attained.
Finally, we demonstrated that the enhancement of quantum synchronization
via continuous measurement and feedback control 
can be achieved with a high probability of success even in the 
stronger quantum regime.

The proposed system can, in principle, be implemented using the current experimental setups; synchronization of a quantum vdP with a harmonic drive can be experimentally implemented using optomechanical systems~\cite{walter2014quantum} or ion traps~\cite{lee2013quantum}, and the feedback control can be implemented by adjusting the frequency of the harmonic drive using the measurement outcomes.
Quantum measurement, an essential feature in quantum systems, helps us resolve the issue of quantum fluctuations that disturb strict quantum synchronization
and is important for the realization and future applications of quantum synchronization in the evolving field of quantum technologies.

{\it Acknowledgments.-}
Numerical simulations are performed using the QuTiP numerical
toolbox \cite{johansson2012qutip,*johansson2013qutip}. 
The authors gratefully thank N. Yamamoto for the stimulating discussions.
We acknowledge JSPS KAKENHI JP17H03279, JP18H03287, JPJSBP120202201, JP20J13778, and JST CREST JP-MJCR1913 for financial support. 

\appendix

\section{Feedback policy}
\begin{figure} [!t]
	\begin{center}
		\includegraphics[width=0.7\hsize,clip]{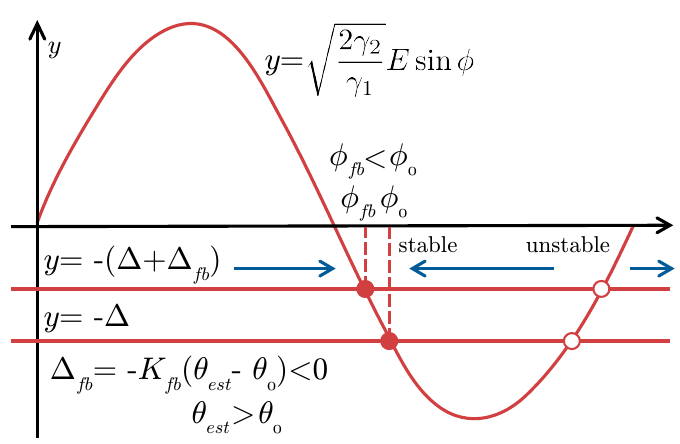}
		\caption{
			Schematic diagram of feedback policy
			for suppressing fluctuations around phase-locking point. 
			Feedback control shifts phase-locking point 
			from $\phi_0$ to $\phi_{fb}$ that is opposite to the
			direction from $\theta_0$ to $\theta_{est}$.
		}
		\label{fig_9}
	\end{center}
\end{figure}

\label{appendix_a}
We discuss the feedback policy for suppressing 
measurement-induced fluctuations around the phase-locking point. 
To understand the core idea of the feedback policy with a simple model,
we consider the system described in Eq.~(\ref{eq:sme}) without 
the linear coupling to the bath, i.e. $\gamma_3 = 0$.
We also assume that the system is in the semiclassical regime
and the oscillator dynamics can be described by a semiclassical stochastic differential equation (SDE) whose deterministic part possesses a stable limit-cycle solution. We can then apply the semiclassical phase reduction~\cite{kato2019semiclassical} to obtain an approximate one-dimensional SDE for the phase variable of the oscillator and use the standard classical methods for the phase equation \cite{winfree2001geometry, kuramoto1984chemical, 
	pikovsky2001synchronization, nakao2016phase, strogatz1994nonlinear}
to analyze synchronization dynamics of the oscillator driven by a periodic forcing.

When the quantum noise is sufficiently weak and the classical limit can be taken,
the deterministic phase equation for the oscillator 
is expressed as (see also the next section) \cite{walter2014quantum,kato2019semiclassical}
\begin{align}
\label{eq:qvdp_dphi_ori}
\frac{d\phi}{dt} &=  
\Delta + \Delta_{fb} + \sqrt{ \frac{2\gamma_{2}}{\gamma_{1}} } E \sin \phi.
\end{align}
When $| \Delta + \Delta_{fb} | \leq \sqrt{ \frac{2\gamma_{2}}{\gamma_{1}} } E$, 
there exists a stable fixed point of Eq.~(\ref{eq:qvdp_dphi_ori}),
which corresponds to the phase-locking point of the system 
with the harmonic driving signal 
under the feedback control, satisfying
\begin{align}
\phi_{fb} = -\arcsin \left( 
\frac{ \Delta + \Delta_{fb}}{E} \sqrt{ \frac{\gamma_{1}}{2\gamma_{2}} } \right).
\end{align}
The fixed point when the feedback control is turned off, i.e., $\Delta_{fb} = 0$, is expressed as
\begin{align}
\phi_0 = - \arcsin \left( 
\frac{ \Delta}{E} \sqrt{ \frac{\gamma_{1}}{2\gamma_{2}} } \right).
\end{align}

Figure~\ref{fig_9} shows a schematic diagram of the feedback policy
for suppressing fluctuations around the phase-locking point.
As shown in Fig.~\ref{fig_9}, when $\theta_{est} > \theta_0$,
the feedback control is $\Delta_{fb} = -K_{fb} (\theta_{est} - \theta_0) < 0$
and hence $ \phi_{fb} < \phi_0$. 
Similarly, when $\theta_{est} < \theta_0$,
we obtain $ \phi_{fb} > \phi_0$.
Therefore, the feedback control shifts the locking phase 
from $\phi_0$ to $\phi_{fb}$, which is opposite to the
direction from $\theta_0$ to $\theta_{est}$,
and is expected to suppress the fluctuations 
of the system around the phase-locking point.

\section{Relationship between the phase diffusion and purity}
\label{appendix_b}
We discuss the relation between the phase diffusion and purity
of the quantum vdP oscillator when the measurement is absent.
We consider the system described in Eq.~(\ref{eq:sme}) without 
the linear coupling to the bath, i.e., $\gamma_3 = 0$;
additionally, we assume that the system is in the semiclassical regime 
and driven by the weak perturbation.
The system can then be 
approximately described by a SDE of the 
phase variable of the oscillator by using the semiclassical phase reduction theory \cite{kato2019semiclassical}.

We introduce the following rescaled quantities:
$\gamma_2 = \sigma \gamma_1 \gamma_2{'},
\Delta + \Delta_{fb} = \gamma_{1} (\Delta' + \Delta'_{fb}), E = \epsilon  \gamma_{1} E{'}/\sqrt{\sigma} ,
dt' = \gamma_1 dt, dW' = \sqrt{\gamma_1}dW$
with dimensionless parameters $\gamma_2{'}$, $\Delta'$, $\Delta'_{fb}$, 
and $E'$ of $\mathcal{O}(1)$.
We set $0 < \sigma \ll 1$ (the system is in the semiclassical regime)
and $0 < \epsilon \ll 1$ (the perturbation is weak).
The corresponding semiclassical phase equation for the quantum system
in Eq.~(\ref{eq:sme}) is then given by \cite{kato2019semiclassical} 

\begin{align}
\label{eq:qvdp_dphi}
d\phi = 
\left( 
\Delta' + \Delta'_{fb}  + \epsilon \sqrt{ 2\gamma'_{2}} E' \sin \phi \right) dt'
+ \sqrt{\sigma D_0}  dW',
\end{align}
with $D_0 = \frac{3 \gamma'_{2}}{2}$.

We first evaluate the phase diffusion of the oscillator
based on the effective diffusion constant 
of Eq.(\ref{eq:qvdp_dphi}) \cite{lifson1962self},

\begin{align}
\label{eq:d_eff}
D_{eff} \propto \frac{1}{(\la \exp( v(\phi)/(\sigma D_0)) \ra_{\phi}  
\la \exp(-v(\phi) /(\sigma D_0)) \ra_{\phi})},
\end{align}
where the potential $v(\phi)$ is given by $v(\phi) = - \int_{\phi_0}^{ \phi} (\Delta' + \Delta'_{fb} + \epsilon \sqrt{ 2\gamma'_{2}} E' \sin \phi' ) d\phi'$ with a reference phase point $\phi_0$
and $\la \cdot \ra_{\phi} = \frac{1}{2\pi} \int_0^{2\pi} ( \cdot ) d\phi$.
When $\sigma$ is sufficiently small, 
using the saddle-point approximation
$\la \exp( v(\phi)/(\sigma D_0)) \ra_{\phi} \approx \exp( v_{max}/(\sigma D_0))$ and 
$\la \exp( -v(\phi)/(\sigma D_0)) \ra_{\phi} \approx \exp( -v_{min}/(\sigma D_0))$,
the effective diffusion constant can be approximated as
\begin{align}
D_{eff} \approx \frac{1}{ \exp( (v_{max}- v_{min})/(\sigma D_0) )},
\end{align}
where $v_{max}$ and $v_{min}$ are the maximum and 
minimum values of the potential $v(\phi)$, respectively
(see \cite{pikovsky2015maximizing} for details), and 
the constant factors are omitted. 

We next evaluate the purity. 
Using semiclassical phase reduction theory \cite{kato2019semiclassical}, 
the density matrix can be approximately reconstructed from the phase equation as 
$
\rho \approx \int_{0}^{2 \pi}d \phi P(\phi)
\left| \alpha_{0}(\phi) \right\rangle \left \langle \alpha_{0}(\phi) \right|,
$
where $\alpha_{0}(\phi) = \sqrt{\frac{1}{2 \sigma \gamma'_{2}}} \exp(i \phi)$ is the system state 
at $\phi$ on the classical limit cycle in the phase space of the P representation 
\cite{gardiner1991quantum}
and $P(\phi)$
is the steady-state probability distribution of the Fokker-Planck equation for the phase variable given by (\cite{pikovsky2001synchronization}, Chapter 9)

\begin{equation}
P(\phi) \propto \int_{0}^{2 \pi}  d\phi' 
\exp \left[ \frac{ 2(v(\phi' + \phi) - v(\phi))}{\sigma D_0 } \right].
\end{equation}
Because the size of the limit cycle is $O(1/\sqrt{\sigma})$, i.e.,
$\alpha_{0}(\phi) = O(1/\sqrt{\sigma})$,
when $\sigma$ is sufficiently small,
the purity can be evaluated 
by using saddle-point approximation as
\begin{align}
& P = \Tr({\rho^{2}}) \approx 
\int_{0}^{2 \pi}  d \phi P(\phi) \int_{0}^{2 \pi} d \phi' P(\phi')
\exp( - \abs{\alpha_{0}(\phi) - \alpha_{0}(\phi')}^2) 
\cr
&\approx 
\int_{0}^{2 \pi} d \phi P^2(\phi)
\approx
\int_{0}^{2 \pi}  d \phi
\left(
\int_{0}^{2 \pi}  d\phi' 
\exp \left[ \frac{ 2(v(\phi' + \phi) - v(\phi))}{\sigma D_0 } \right]
\right)^2
\cr
& \approx
\int_{0}^{2 \pi}  d \phi
\left(
\exp \left[ \frac{2(v_{max} - v(\phi))}{\sigma D_0 } \right]
\right)^2
\approx
\exp \left[ \frac{4(v_{max} - v_{min})}{\sigma D_0 } \right],
\nonumber
\end{align}
where the constant factors are omitted.
The effective diffusion constant $D_{eff}$ 
can then be approximately represented as
\begin{align}
D_{eff} \propto \frac{1}{ (P )^{1/4} },
\end{align}
which indicates that a higher purity results in a smaller phase diffusion
of the oscillator.

\end{document}